\documentclass[12pt]{iopart}
\usepackage{iopams}  
\usepackage{graphicx}

\newcommand{\beeq}{\begin{equation}}
\newcommand{\eneq}{\end{equation}}
\newcommand{\beeqa}{\begin{eqnarray}}
\newcommand{\eneqa}{\end{eqnarray}}
\newcommand{\cH}{\mathcal H}
\newcommand{\cR}{{\cal R}}
\newcommand{\cA}{\mathcal A}
\newcommand{\dd}{\partial}

\begin{document}

\title[Primordial magnetic fields from the non-adiabatic
fluctuations]
{Primordial magnetic fields generated by the non-adiabatic
fluctuations at pre-recombination era}

\author{Satoshi Maeda$^1$, Keitaro Takahashi$^2$ and Kiyotomo Ichiki$^3$}

\address{$^1$Department of Physics, Kyoto University, Kyoto 606-8502, Japan}
\address{$^2$Department of Physics, Kumamoto University, Kumamoto 860-8555, Japan}
\address{$^3$Department of Physics, Nagoya University, Nagoya 464-8602, Japan}
\eads{\mailto{smaeda"at"tap.scphys.kyoto-u.ac.jp}, \mailto{keitaro"at"sci.kumamoto-u.ac.jp},
     \mailto{ichiki"at"a.phys.nagoya-u.ac.jp}}

\date{\today}
\begin{abstract}
 In the pre-recombination era, cosmological density fluctuations can
 naturally generate magnetic fields through Thomson scatterings. In
 previous studies, only the magnetic field generation from the initially-
 adiabatic fluctuations has been considered.  Here we investigate the
 generation of cosmological magnetic fields sourced by the primordial
 non-adiabatic fluctuations based on the cosmological perturbation
 theory, using the tight-coupling approximations between photon and
 baryon fluids.  It is found that the magnetic fields from the
 non-adiabatic fluctuations can arise at the first-order expansion of
 the tight coupling approximation. This result is in contrast to the
 case of adiabatic initial fluctuations, where the magnetic fields can
 be generated only at the second-order.  In a general case where the
 primordial density perturbations contain small non-adiabatic
 fluctuations on the top of the dominant adiabatic ones, we show that
 the leading source of magnetic fields is given by the second-order
 coupling of the adiabatic and non-adiabatic fluctuations.  We calculate
 the power spectrum of the generated magnetic fields when the
 non-adiabatic fluctuations have a blue power spectrum, which has been
 suggested by recent cosmological observations.
\end{abstract}
\vspace{3mm}
\begin{flushleft}
\end{flushleft}

\maketitle

\section{Introduction}
The magnetic fields in galaxies and in clusters of
galaxies are observed and the typical strength is about 1$\mu$Gauss.
The origin of these fields is not clear yet
\cite{Grasso:2000wj,Widrow:2002ud,Giovannini:2003yn,Kandus:2010nw}.
This is one of the significant problem in the modern cosmology.
It has been considered that very weak large-scale magnetic fields generated 
in the early universe are amplified by the dynamo mechanism after galaxy
formation.
If the large-scale fields exist in the early universe, the cosmic
microwave background or the large-scale structure formation are affected.
We have the upper-limit of the large-scale magnetic field
, $<10^{-9}$Gauss, from CMB observations
\cite{Yamazaki:2008gr, Yamazaki:2010nf, Kahniashvili:2010wm,
Tashiro:2009hx,Paoletti:2010rx, Schleicher:2011jj}.
On the one hand, astrophysical activities are considered to vanish 
information of the primordial magnetic fields near astrophysical objects
but not the fields in the void regions since the objects rarely exist in void.
There is an attempt to constrain these fields using the delayed or
extended $\gamma$-ray emissions from distant 
high-energy astrophysical objects, for example, quasars.
We have the lower limit $\gtrsim 10^{-17}$ Gauss in the void 
under some assumptions
\cite{Ando:2010rb,Essey:2010nd,Neronov:1900zz,Dolag:2010ni,Takahashi:2011ac}.

Many mechanisms generating primordial magnetic fields are proposed,
such as the generation from the second-order density perturbation 
at the pre-recombination era
\cite{Berezhiani:2003ik,Matarrese:2004kq,Takahashi:2005nd,Kobayashi:2007wd,Ichiki:2007hu,Maeda:2008dv,Fenu:2010kh}.
In this era, protons and electrons move interacting with photons by the
Thomson scattering.  As the mass of electrons is much smaller than
protons, the cross section of the Thomson scattering for electrons is
larger and it causes the electric fields.  The curly components of these
electric fields generate the magnetic fields at the pre-recombination era.

The primordial fluctuations are generated in the very early universe,
for example, at the time of inflation.
They evolve into the anisotropies of the cosmic microwave
background (CMB) and act as the seed of the large-scale structure.
In a simplest single-field inflation model, only adiabatic fluctuations 
are produced. However in other models such as a multi-field inflation model 
or a curvaton model, non-adiabatic fluctuations are also produced
since a new degree of freedom is added in the system \cite{Bassett:2005xm}.
Actually, adiabatic fluctuations are known as the main component 
of the fluctuations from the CMB observations.  However,
the non-adiabatic fluctuations can still exist as a sub-dominant component.
For example, in refs. \cite{Sollom:2009vd,Li:2010yb}, it is indicated that 
the power spectrum of non-adiabatic fluctuations may have 
a very blue spectrum. On the one hand, the possibility of generating
vorticity  by the non-adiabatic fluctuations is also studied recently
\cite{Christopherson:2009bt,Christopherson:2010ek,Pitrou:2010ai}.

If the non-adiabatic fluctuations exist, magnetic fields are also
generated as is expected from the fact that magnetic fields follow similar
evolution equations as vorticities.
In previous studies,
initially-adiabatic fluctuations are considered for the generation of
magnetic fields.  Even if one starts from the adiabatic initial conditions,
the non-adiabatic fluctuations such as differences in motion between
photons and baryons are generated and they lead to the generations
of magnetic fields\footnote{
There is another spcial mechanism of the magnetic field generation from non-adiabatic
fluctuations relating cosmic strings 
\cite{Brandenberger:1998ew,Gwyn:2008fe}.}. In this case, however, it has been shown that
the fields can not be generated at the first order of the tight coupling
approximation (TCA) for the Thomson scattering and we need to consider
the second order \cite{Kobayashi:2007wd,Maeda:2008dv}. This means
that the amplitude of the generated magnetic fields is suppressed
by a tight coupling parameter.

In this paper, we will show that the fields can be generated at
the first-order expansion of the TCA if the primordial non-adiabatic
fluctuations exist, and show that the leading source of the fields is
given by the cross term of the primordial adiabatic and
primordial non-adiabatic fluctuations. 
Hence, there is a possibility that the strength of the primordial
magnetic fields is stronger than that of the fields only from adiabatic 
fluctuations if the non-adiabatic fluctuations have a blue spectrum.
Therefore it is important to examine the magnetic fields generated 
by the non-adiabatic fluctuations.  The aim of this paper is thus to
show explicitly that the primordial non-adiabatic fluctuations can 
generate the magnetic fields at the first-order expansion of the TCA and
to obtain and evaluate the power spectrum of the generated magnetic fields.

The plan of our paper is the following.
We show that the non-adiabatic fluctuations generate the magnetic fields
at the first-order expansion in the TCA in the next section.
The calculation is almost the same as the process in
ref.\cite{Maeda:2008dv}, but we now include the non-adiabatic
fluctuations at initial conditions.
In section 3, we derive the power spectrum of the generated magnetic
fields.
We present the results in section 4 and summarize in section 5.

\section{Generation of the magnetic fields}
\subsection{Evolution equation of the magnetic fields}
We need to consider the second-order cosmological perturbation theory 
because no magnetic fields are generated from scalar-type perturbation
at the first order.
First we define the metric used in this paper.
We consider the following metric in the Poisson gauge 
under the flat-Friedmann spacetime:
\begin{equation}
ds^2 = a^2(\eta )\left[-\left(1+2\phi\right)d\eta^2 
       + 2\chi_id\eta dx^i
+  \left( 1-2\cR \right)\delta _{ij} dx^idx^j\right],
\label{metric}
\end{equation}
where $a$ and $d\eta=dt/a$ is scale factor and conformal time.
The each metric perturbation is expanded as 
\begin{equation}
\phi=\phi^{(1)} + \phi^{(2)},~\cR=\cR^{(1)}+\cR^{(2)},~
 \chi_i=\chi_i^{(2)},
\end{equation}
where Arabic number represents the order of the cosmological perturbation.

The basic equations are the equations of motion (EoM) of the matter and
the induction equation of the magnetic fields.
In pre-recombination era, photon($\gamma$), proton($p$) and 
electron($e$) suffuse the universe and interact with each other.
Since we assume the charge neutrality, the number density for charged particles
is the same $n = n_p\simeq n_e$.
Then the EoMs for each spices are
\begin{eqnarray}
\nabla_{\nu}T_{(\gamma )i}{}^{\nu} &=& \kappa^{\gamma p}_i + \kappa^{\gamma e}_i , \label{ph}\\
\nabla_{\nu}T_{(p)i}{}^{\nu} &=& enE_i + \kappa^{pe}_i + \kappa^{p\gamma }_i , \label{pr}\\
\nabla_{\nu}T_{(e)i}{}^{\nu} &=& -enE_i  + \kappa^{ep}_i + \kappa^{e\gamma }_i , \label{el}
\end{eqnarray}
where $E_i$ and $\kappa^{ab}_i$ in the right hand side are the electric
fields and collision terms
\cite{Ichiki:2007hu,Bartolo:2006cu,Pitrou:2008ut}.
The Thomson scattering between photons and charged particles is 
\begin{eqnarray}
\kappa^{\gamma e}_i &= -\kappa^{e\gamma}_i 
= -\frac{4}{3}\sigma_T n\rho_\gamma (1-2\cR)\left[
\left(v_{(\gamma) i} -v_{(e)i}\right)-\frac{1}{4}v_{(e)j}\Pi_{\gamma~i}^{~j}
\right] ,
\label{tr1}
\\
\kappa^{\gamma p}_i &= -\kappa^{p\gamma}_i = 
-\frac{m_e^2}{m_p^2}\frac{4}{3}\sigma_T n\rho_\gamma (1-2\cR)\left[
\left(v_{(\gamma) i} - v_{(p)i}\right)
-\frac{1}{4}v_{(p)j}\Pi_{\gamma~i}^{~j}
\right],
\label{tr2}
\end{eqnarray}
where $\sigma_T$ and $\Pi_{\gamma,i}^{~j}$ are the cross section of the 
Thomson scattering and anisotropic stress of photons.
The Coulomb scattering between charged particles is
\begin{equation}
\kappa^{pe}_i = -\kappa^{ep}_i = -e^2n^2\eta_C 
(1-2\cR)\left(v_{(p) i} -v_{(e)i}\right),
\end{equation}
where $\eta_C$ is the electric resistivity.
We also use the induction equation of the magnetic fields (\ref{appind})
\begin{equation}
\left(a^3B^i\right)' 
=-\epsilon^{ijk}\dd_j\left[a\left(1+\phi\right)E_k\right]
      -\epsilon^{ijk}\left(av_jE_k\right)',
\label{indB}
\end{equation}
where the prime represents the derivative with conformal time $\eta$.

Subtracting two EoMs for protons and electrons leads to 
``the Ohm's law''.
In our case, there are two velocity differences, between photon and
baryon $\delta v_{\gamma b}=v_{\gamma}-v_b$ and between proton and
electron $\delta v_{pe}=v_p-v_e$ where $v_b$ is the baryon's velocity of the
center of mass.
We assume that proton and electron are strongly coupled because coulomb
interaction is strong.
It results that the velocity difference between protons and electrons is
very small compared with one between photons and baryons.
Even if the electric current does not exist, the magnetic field can be generated.
As a result, we obtain ``the Ohm's law'' 
\begin{equation}
E_i = \frac{1-\beta^3}{1+\beta}\frac{4\sigma _T}{3e}a\rho_\gamma
       (1-2\cR)\Delta v_{(\gamma b)i}
,\label{ohmslaw}
\end{equation} 
where we define 
\begin{equation}
\Delta v_{(\gamma b)i}:= 
   \delta v_{(\gamma b)i} - \frac{1}{4}v_{(b)j}\Pi_{\gamma~i}^{~j},
\end{equation}
$\beta=m_e/m_p$ as the ratio between the mass 
of proton and electron and neglect $\delta v_{(pe)i}$.
If the mass of proton and electron is same, $\beta=1$, the electric
fields are not generated.
Eq.(\ref{ohmslaw}) corresponds to the equation added anisotropic stresses 
to eq.(27) in ref.\cite{Maeda:2008dv}.

Substituting eq.(\ref{ohmslaw}) into the induction equation eq.(\ref{indB}),
we obtain the evolution equation of the magnetic fields
\begin{eqnarray}
&&\left(a^3B^i\right)' 
= -\frac{1-\beta^3}{1+\beta}\frac{4\sigma _T}{3e}
    \epsilon^{ijk}a^2
\times\nonumber\\ &&\quad\times
\left[
    \dd_j\left(\rho_\gamma\Delta v_{(\gamma b)k}\right) 
     + \rho_\gamma\dd_j(\phi-2\cR)\Delta v_{(\gamma b)k}
      +\frac{1}{a^2}\left(\rho_\gamma v_ja^2\Delta v_{(\gamma b)k}\right)'
\right].\label{Mag1}
\end{eqnarray}
The equation means that the sources generating the magnetic field are 
the velocity difference between photons and baryons and 
anisotropic stress of photons.

\subsection{Analytic formula of the velocity difference $\Delta v_{(\gamma b)}$}
We use the tight coupling approximation (TCA) for the Thomson scattering 
so as to represent $\Delta v_{(\gamma b)i}$ by the well-known
perturbative quantities.
When the dynamical timescale is longer than the timescale 
of the Thomson scattering, we can expand the physical quantities under the TCA.
At zeroth order in the TCA, all fluid components move together 
with the same velocity $v_i$.
We set the photon frame as the ``background'' in the TCA. 
In other words, physical quantity for zeroth-order TCA is the
same as that of photon.
The velocity for baryons is expanded as
\begin{eqnarray}
v_{(b)i} = v_i+v_{(b)i}^{(I)}+v_{(b)i}^{(II)}+\cdots,
\end{eqnarray}
where $v_i$ is the common velocity of photons and baryons 
in the tight coupling limit and also the photon's velocity.
Since various physical quantities such as
$\delta v_{(\gamma b)i}$, $\Pi_{\gamma~i}^{~j}$ 
are expanded 
\begin{eqnarray}
&& \delta v_{(\gamma b)i} = \delta v_{(\gamma b)i}^{(I)} 
 + \delta v_{(\gamma b)i}^{(II)}+\cdots,\\
&&\Pi_{\gamma~i}^{~j}=\Pi_{\gamma~i}^{(I)j}+\Pi_{\gamma~i}^{(II)j}+\cdots,
\end{eqnarray}
the quantity $\Delta v_{(\gamma b)i}$ in the eq.(\ref{Mag1}) is expressed
\begin{eqnarray}
 \Delta v_{(\gamma b)i} = \Delta v_{(\gamma b)i}^{(I)} 
 + \Delta v_{(\gamma b)i}^{(II)}+\cdots.
\end{eqnarray}

The equation which determines the velocity difference between 
baryon and photon is provided from the subtraction of the EoM for
photon from baryon
\begin{eqnarray}
& & 
\frac{\rho_\gamma'}{\rho_\gamma}\left(v_{(\gamma)i}+\chi_i\right) 
     -\frac{n'}{n}\left(v_{(b)i}+\chi_i\right) 
 +\left(\delta v_{(\gamma b)i}\right)' +4\cH\delta v_{(\gamma b)i} 
\nonumber\\&&\qquad
-\left(\phi+2\cR\right)\left[\frac{\rho_\gamma'}{\rho_\gamma}v_{(\gamma)i}
-\frac{n'}{n}v_{(b)i}
  +\left(\delta v_{(\gamma b)i}\right)'+4\cH\delta v_{(\gamma b)i}\right]
\nonumber\\&&\qquad
- 5\cR'\delta v_{(\gamma b)i}
+ \frac{1}{4}\frac{\dd_i\rho_\gamma}{\rho_\gamma}
+\dd_j\left(v_{(\gamma)i}v_{(\gamma)}^j\right)
 -\frac{1}{1+\beta}\dd_j\left(v_{(p)i}v_{(p)}^j+\beta v_{(e)i}v_{(e)}^j\right) 
\nonumber\\
&&=-\alpha (1-2\cR)\Delta v_{(\gamma b)i}
\label{evo-v},
\end{eqnarray} 
where $\alpha$ is defined as 
\begin{equation}
\alpha := \frac{1+\beta^2}{1+\beta}(1+R)\frac{4a\sigma_T \rho_\gamma }{3m_p}, 
\end{equation}
with $R:= (3m_p(1+\beta)n)/(4\rho_\gamma) $.
From the next section, we solve $\Delta v_{(\gamma b)i}$ under 
the above expansion of the tight coupling approximation.

\subsubsection{Background of the TCA}
Firstly we note the background in the TCA and define 
the non-adiabatic fluctuations.
In the zeroth-order TCA and the first-order cosmological perturbation, 
the energy conservations for photon and baryon, 
\begin{eqnarray}
\left(\frac{\delta\bar{\rho}_\gamma^{(1)}}{\bar{\rho}^{(0)}_\gamma}\right)'
= 4\left(\cR^{(1)}\right)' -\frac{4}{3}\dd_i v^{i(1)} ,\quad
 \left(\frac{\delta \bar{n}^{(1)}}{\bar{n}^{(0)}}\right)'
= 3\left(\cR^{(1)}\right)'-\dd_i v^{i(1)} 
\label{ene1},
\end{eqnarray}
imply that
\begin{equation}
\left(\frac{\delta\bar{\rho}_\gamma^{(1)}}{\bar{\rho}^{(0)}_\gamma}\right)'
=\frac{4}{3}\left(\frac{\delta \bar{n}^{(1)}}{\bar{n}^{(0)}}\right)', 
\end{equation}
where the bar represents the quantity in the zeroth-order TCA.
In the case of primordial non-adiabatic fluctuations, there is 
a difference between the fluctuations of photons and baryons 
as an initial condition which means
\begin{equation}
\frac{\delta \bar{n}^{(1)}}{\bar{n}^{(0)}}
= \frac{3}{4}\frac{\delta\bar{\rho}_\gamma^{(1)}}{\bar{\rho}^{(0)}_\gamma}
 + \bar{C}(\vec x).
\label{cdef}
\end{equation}
In the above equation, $C(\vec x)$ corresponds to the non-adiabatic
fluctuation, which is usually called baryon isocurvature mode.
This indicates that each fluctuations move 
with the common velocity keeping the difference in density contrast 
which is caused by the non-adiabatic fluctuations.
In the adiabatic initial condition, this term becomes zero, so photons and
baryons move together, as one fluid.
As we shall show later $C(\vec{x})$ plays a central role in generating
magnetic fields.

\subsubsection{First-order cosmological perturbation}
The first-order cosmological perturbation is the most simple calculation.
In this order,  eq.~(\ref{evo-v}) presents
\begin{equation}
\left[\frac{\left(\bar{\rho}_\gamma^{(0)}\right)'}{\bar{\rho}_\gamma^{(0)}}
        -\frac{\left(\bar{n}^{(0)}\right)'}{\bar{n}^{(0)}}\right]v_i^{(1)}
+\frac{1}{4}\frac{\dd_i \delta\bar{\rho}_\gamma^{(1)}}{\bar{\rho}_\gamma^{(0)}}
= -\bar{\alpha}^{(0)}\Delta v_{(\gamma b)i}^{(I,1)},
\end{equation}
and so we obtain
\begin{equation}
\Delta v_{(\gamma b)i}^{(I,1)} 
= \frac{1}{\bar{\alpha}^{(0)}}\left[\cH v_i^{(1)}
    -\frac{1}{4}\frac{\dd_i
    \delta\bar{\rho}_\gamma^{(1)}}{\bar{\rho}_\gamma^{(0)}}\right], 
\label{dv11}
\end{equation}
where  we use background equations and define
\begin{equation}
\bar \alpha^{(0)}=\frac{(1+\beta^2)(1+\bar R^{(0)})}{1+\beta}
  \frac{4a\sigma_T\bar{\rho}_\gamma^{(0)}}{3m_p} .\label{baralpha0} 
\end{equation}
In this order, the non-adiabatic fluctuations do not contribute 
because the density fluctuations of baryon do not appear.

\subsubsection{Second-order cosmological perturbation}
The effect of the non-adiabatic fluctuations and anisotropic stress
appear firstly in this order.
Equation (\ref{evo-v}) becomes
\begin{eqnarray}
&&\Biggl[\frac{\left(\delta\bar{\rho}_\gamma^{(1)}\right)'}{\bar{\rho}_\gamma^{(0)}}
 -\frac{\delta\bar{\rho}_\gamma^{(1)}}{\bar{\rho}_\gamma^{(0)}}
  \frac{\left(\bar{\rho}_\gamma^{(0)}\right)'}{\bar{\rho}_\gamma^{(0)}}
 -\left\{\frac{\left(\delta\bar{n}^{(1)}\right)'}{\bar{n}^{(0)}}
          -\frac{\delta\bar{n}^{(1)}}{\bar{n}^{(0)}}
  \frac{\left(\bar{n}^{(0)}\right)'}{\bar{n}^{(0)}}\right\}
  \nonumber\\&&\qquad
 -\left(\phi^{(1)}+2\cR^{(1)}\right)
   \left\{\frac{\left(\bar{\rho}_\gamma^{(0)}\right)'}{\bar{\rho}_\gamma^{(0)}}
        -\frac{\left(\bar{n}^{(0)}\right)'}{\bar{n}^{(0)}}\right\}
\Biggr]v_i^{(1)}
\nonumber\\&&\qquad
+\left[\frac{\left(\bar{\rho}_\gamma^{(0)}\right)'}{\bar{\rho}_\gamma^{(0)}}
        -\frac{\left(\bar{n}^{(0)}\right)'}{\bar{n}^{(0)}}\right]
\left(v_i^{(2)}+\chi^{(2)}_i\right)
-\frac{1}{4}\left[
\frac{\delta\bar{\rho}_\gamma^{(1)}}{\bar{\rho}_\gamma^{(0)}}
   \frac{\dd_i \delta\bar{\rho}_\gamma^{(1)}}{\bar{\rho}_\gamma^{(0)}}
-\frac{\dd_i \delta\bar{\rho}_\gamma^{(2)}}{\bar{\rho}_\gamma^{(0)}}
\right]
\nonumber\\&&
=- \bar{\alpha}^{(0)}\Delta v_{(\gamma b)i}^{(I,2)}
 + 2\cR^{(1)}\bar{\alpha}^{(0)}\Delta v_{(\gamma b)i}^{(I,1)}
 - \bar{\alpha}^{(1)}\Delta v_{(\gamma b)i}^{(I,1)}
\label{Cos2nd}
,
\end{eqnarray}
where 
\begin{eqnarray}
\frac{\bar{\alpha}^{(1)}}{\bar{\alpha}^{(0)}}
&=&
\frac{1}{\Bigl(1+\bar{R}^{(0)}\Bigr)}
  \Biggl[\bar{R}^{(0)}\frac{\delta\bar{\rho}_\gamma^{(1)}}{\bar{\rho}_\gamma^{(0)}}
 +\bar{R}^{(0)}\frac{\delta \bar{n}^{(1)}}{\bar{n}^{(0)}}
 \Biggr]
\nonumber\\
&=&\frac{4+3\bar{R}^{(0)}}{4\Bigl(1+\bar{R}^{(0)}\Bigr)}
 \frac{\delta\bar{\rho}_\gamma^{(1)}}{\bar{\rho}_\gamma^{(0)}}
 +\frac{\bar{R}^{(0)}}{1+\bar{R}^{(0)}}\bar{C}^{(1)}
:= 
\left. \frac{\bar{\alpha}^{(1)}}{\bar{\alpha}^{(0)}} \right|_{\rm ad}
+\frac{\bar{R}^{(0)}}{1+\bar{R}^{(0)}}\bar{C}^{(1)}
.
\end{eqnarray}
Baryon density perturbation, $\delta \bar{n}^{(1)}$, is 
in both sides of eq.(\ref{Cos2nd}), however,
the non-adiabatic term in the left-hand side vanishes 
because the terms in the curly brackets become 
$(\delta \bar{n}^{(1)}/\bar{n}^{(0)})'$ and 
non-adiabatic fluctuations are independent of time at this
order from eq.(\ref{cdef}).
This can be solved for $\Delta v_{(\gamma b)i}^{(I,2)}$ to give
\begin{eqnarray}
\Delta v_{(\gamma b)i}^{(I,2)} 
&=& \frac{1}{\bar{\alpha}^{(0)}}\Bigg[
    \cH\left(v_i^{(2)}+\chi_i^{(2)}
  - \left. \frac{\bar{\alpha}^{(1)}}{\bar{\alpha}^{(0)}} \right|_{\rm
  ad}v_i^{(1)}\right) 
  -\left\{\left(\cR^{(1)}\right)' -\frac{1}{3}\dd_\ell v^{\ell(1)}\right\}v_i^{(1)}
\nonumber \\ &&
 -\cH\phi^{(1)}v_i^{(1)}
   -\frac{1}{4}\left\{
    \frac{\dd_i \delta\bar{\rho}_\gamma^{(2)}}{\bar{\rho}_\gamma^{(0)}}  
   -\left(\frac{\delta\bar{\rho}_\gamma^{(1)}}{\bar{\rho}_\gamma^{(0)}}
     +\left.\frac{\bar{\alpha}^{(1)}}{\bar{\alpha}^{(0)}}\right|_{\rm ad}\right)
       \frac{\dd_i \delta\bar{\rho}_\gamma^{(1)}}{\bar{\rho}_\gamma^{(0)}}
\right\}
  - \frac{\cR^{(1)}}{2}\frac{\dd_i
    \delta\bar{\rho}_\gamma^{(1)}}{\bar{\rho}_\gamma^{(0)}}
\Bigg] 
\nonumber\\
&& -
 \frac{\bar{R}^{(0)}}{1+\bar{R}^{(0)}}\bar{C}^{(1)}\Delta v_{(\gamma b)i}^{(I,1)} 
.
\label{dv12}
\end{eqnarray}
The last term is the contribution of the non-adiabatic fluctuations 
and the anisotropic stress.

\subsection{Source term of the magnetic fields}
Here we show the generation of the magnetic fields from the non-adiabatic 
fluctuations.
We substitute eqs.(\ref{dv11}) and (\ref{dv12}) into eq.(\ref{Mag1}) and
obtain the evolution equation of the magnetic fields
\begin{eqnarray}
&&\left(a^3B^i\right)' 
=
\frac{1-\beta^3}{1+\beta}\frac{4\sigma _T}{3e}
    a^2\bar{\rho}^{(0)}_{\gamma}
\epsilon^{ijk}\Bigg[\partial_j \Delta v_{(\gamma b)k}^{(I,2)}
 + \frac{1}{\bar \rho_\gamma^{(0)}} \partial_j 
\left(\delta \bar \rho_\gamma^{(1)} \Delta v_{(\gamma b)k}^{(I,1)}\right)
\nonumber\\&&\qquad\qquad\qquad
+ \partial_j (\phi^{(1)} - 2\cR^{(1)})\Delta v_{(\gamma b)k}^{(I,1)} 
+ \frac{1}{a^2 \bar \rho_\gamma^{(0)}}
   \left(a^2 \bar{\rho}_\gamma^{(0)}v_j^{(1)} 
   \Delta v_{(\gamma b)k}^{(I,1)}\right)' \Bigg]
\nonumber\\&&\quad
= \frac{1-\beta^3}{1+\beta}\frac{4\sigma _T}{3e}
    a^2\bar{\rho}^{(0)}_{\gamma}\epsilon^{ijk}\left[
\frac{2a^2\cH}{\bar \alpha^{(0)}}\omega^{(2)i}
+
 \frac{\bar{R}^{(0)}}{1+\bar{R}^{(0)}}\epsilon^{ijk}
 \dd_j\bar{C}^{(1)}\Delta v_{(\gamma b)k}^{(I,1)}
\right],\label{Mag2}
\end{eqnarray}
where we use
\begin{eqnarray}
\epsilon^{ijk}\dd_j\Delta v_{(\gamma b)k}^{(I,2)}
&=&-\frac{2a^2\cH}{\bar \alpha^{(0)}}\omega^{(2)i}
+ 2\epsilon^{ijk}\dd_j\cR^{(1)}\Delta v_{(\gamma b)k}^{(I,1)}
- \epsilon^{ijk}v_j^{(1)}\left[(\Delta v_{(\gamma b)k}^{(I,1)})'
   +\cH\Delta v_{(\gamma b)k}^{(I,1)}\right]
\nonumber\\
&&-
 \frac{\bar{R}^{(0)}}{1+\bar{R}^{(0)}}\epsilon^{ijk}
 \dd_j\bar{C}^{(1)}\Delta v_{(\gamma b)k}^{(I,1)},
\label{eq71}
\end{eqnarray}
with photon's vorticity $\omega^{(2)i}$ which is defined 
in eq.(\ref{appom}).
Equation (\ref{Mag2}) tells us that the non-adiabatic fluctuations 
contribute as the source at the first-order TCA.
This results from the fact that the direction of the velocity difference 
between photon and baryon is different from the gradient 
of the non-adiabatic fluctuations.

%

The non-adiabatic fluctuations generate also the vorticity in eq.(\ref{Mag2}).
We take curl of a total momentum conservation at first-order TCA and 
obtain 
\begin{eqnarray}
&&\left(a^2\omega^{(2)i}\right)'
+ \frac{\cH\bar{R}^{(0)}}{1+\bar{R}^{(0)}}a^2\omega^{(2)i}
= 
 \frac{\bar{R}^{(0)}}{2(1+\bar{R}^{(0)})^2}\bar\alpha^{(0)}
  \epsilon^{ijk}\dd_j\bar{C}^{(1)} \Delta v^{(I,1)}_{(\gamma b)k}.
\label{voleq}
\end{eqnarray}
If the non-adiabatic fluctuations do not exist, the vorticity
decays with time and the magnetic fields can not be generated at this order.
That agree with refs.\cite{Kobayashi:2007wd,Maeda:2008dv}
\footnote{Eqs.(\ref{Mag2}) and (\ref{voleq}) look different 
from eq.(D7) in ref.\cite{Fenu:2010kh}.
However  this comes from the difference of the frame.
Our frame is the same as the frame in refs.\cite{Kobayashi:2007wd,Maeda:2008dv}.
There are the details in Appendix D in ref.\cite{Fenu:2010kh}.
}.

\section{Power spectrum of the generated magnetic fields}
\subsection{Fourier transformation of the source term}
We will derive the power spectrum of the magnetic fields generated by
the non-adiabatic fluctuations.
The source term in eq.(\ref{Mag2}) is
\begin{eqnarray}
S^i_B &=&
\frac{2a^2\cH}{\bar \alpha^{(0)}} \omega^{(2)i}
+
 \frac{\bar{R}^{(0)}}{1+\bar{R}^{(0)}}\epsilon^{ijk}
 \dd_j\bar{C}^{(1)}\Delta v_{(\gamma b)k}^{(I,1)}
\label{s4}
.
\end{eqnarray}
Since we consider first-order scalar perturbations, 
the velocity is given by a gradient of a scalar function, $v$.
Defining 
\begin{equation}
v_i^{(1)} (x^i,\eta) := \dd_i v^{(1)}
,~~
\delta^{(1)}(x^i,\eta) := \frac{\delta\bar{\rho}_\gamma^{(1)}}
 {\bar{\rho}_{\gamma}^{(0)}}
\end{equation}
and
\begin{equation}
\delta V^{(1)} := \cH v^{(1)} - \frac{1}{4}\delta^{(1)}
\label{s},
\end{equation}
then we can write $\Delta v_{(\gamma b)i}^{(I,1)}$ in Eq.(\ref{s4}) as
$ \Delta v_{(\gamma b)i}^{(I,1)} = \dd_i \delta V^{(1)}/\bar{\alpha}^{(0)} $.
The Fourier transformation of eq.(\ref{s4}) is 
\begin{eqnarray}
\hat{S}^i_B&=&
\frac{2a^2\cH}{\bar \alpha^{(0)}}
\hat{\omega}^{(2)i}
+ \frac{\bar{R}^{(0)}}{1+\bar{R}^{(0)}}\frac{1}{(2\pi)^{3/2}}
\frac{1}{\bar \alpha^{(0)}}\int d^3p(\vec{k}\times\vec{p})^i
\hat{C}_{\vec{p}}^{(1)}\hat{\delta V}_{\vec{k}-\vec{p}}^{(1)},
\label{s2}
\end{eqnarray}
where a hat represents the Fourier transformation.
We solve eq.(\ref{voleq}) for the vorticity to obtain
\begin{eqnarray}
a^2\hat \omega^{(2)}
=
\frac{1}{(2\pi)^{3/2}}\frac{1}{1+\bar R^{(0)}}
\int d\eta' 
\frac{\bar{R}^{(0)}}{2(1+\bar{R}^{(0)})}\int d^3p
  (\vec{k}\times\vec{p})^i \hat{C}_{\vec p}^{(1)} 
\hat{\delta V}^{(1)}_{\vec{k}-\vec{p}}.
\label{homega}
\end{eqnarray}
Therefore the Fourier transformation of the magnetic fields 
is
\begin{eqnarray}
\hat{B}^i=
\frac{1-\beta^3}{1+\beta}\frac{\sigma_T\bar\rho_{\gamma0}^{(0)}}{ea^3}
\int \frac{d\eta'}{a^{-2}}\hat{S}^i_B
=
\frac{1-\beta^3}{1+\beta}\frac{\sigma_T\bar\rho_{\gamma0}^{(0)}}{ea^3}
\int^{\eta}_{0}\frac{d\eta'}{a^{-2}}\left( \hat{\Omega}^i + \hat{S}^i\right)
\label{magmag}.
\end{eqnarray}
where $\bar \rho_{\gamma0}^{(0)}$ is the present photon energy density and
\begin{eqnarray}
\hat{S}^i &:=& 
\frac{1}{(2\pi)^{3/2}}\frac{\bar{R}^{(0)}}{1+\bar{R}^{(0)}}
\frac{1}{\bar \alpha^{(0)}}\int d^3p(\vec{k}\times\vec{p})^i
\hat{C}_{\vec{p}}^{(1)}\hat{\delta V}_{\vec{k}-\vec{p}}^{(1)}
\label{sou1}~,\\
\hat{\Omega}^i &:=& 
\frac{2a^2\cH}{\bar \alpha^{(0)}}
\hat{\omega}^{(2)i}
\nonumber\\
&=&\frac{1}{(2\pi)^{3/2}}
\frac{\cH}{\bar \alpha^{(0)}}\frac{1}{1+\bar R^{(0)}}
\int^\eta_0 d\eta'
\frac{\bar{R}^{(0)}}{1+\bar{R}^{(0)}}\int d^3p
  (\vec{k}\times\vec{p})^i \hat{C}_{\vec p}^{(1)} 
\hat{\delta V}^{(1)}_{\vec{k}-\vec{p}}~
\label{om1}
.
\end{eqnarray}

\subsection{Evolution of the fluctuations}
In order to evaluate the source terms, we need to know the evolution 
of $\hat{\delta V}_{\vec k}$.
In the linear order, $\hat{\delta V}_{\vec k}$ is given by 
the sum of the adiabatic and non-adiabatic modes.
Though both the fluctuations should be responsible for the generation of magnetic fields,
we, hereafter, consider only the adiabatic mode in calculating 
$\hat{\delta V}_{\vec k}$. This is 
because the non-adiabatic fluctuations are expected to be smaller than the adiabatic
fluctuation from observations.
Accordingly, the equations (\ref{s2}) and (\ref{homega}) tell that the source of
the magnetic field is the cross coupling of the adiabatic fluctuation
($\hat{\delta V}_{\vec k}$) and the non-adiabatic fluctuation
($\hat{C}_{\vec k}$).

We consider the radiation-dominated era when the Thomson scattering is
strong and the tight coupling approximation is good enough.
Then we can neglect matter perturbations and anisotropic stress in
linearized Einstein equations, namely $\hat{\delta}_{\vec{k}}^{(1)}= \hat{\delta}_{\gamma,\vec{k}}^{(1)}
+ \hat{\delta}_{m,\vec{k}}^{(1)}(\bar\rho_m^{(0)}/\bar\rho_\gamma^{(0)})
\simeq \hat{\delta}_{\gamma,\vec{k}}^{(1)}$
and $\Pi_{\gamma}=0$.
The linearized Einstein equations become
\begin{eqnarray}
&&3\cH\left((\hat{\cR}_{\vec{k}}^{(1)})' + \cH\hat{\cR}_{\vec{k}}^{(1)}\right)
 + k^2\hat{\cR}_{\vec{k}}^{(1)} 
= -4\pi Ga^2\bar{\rho}_\gamma^{(0)}\hat{\delta}_{\vec{k}}^{(1)} 
~\label{e00},\\
&&(\hat{\cR}_{\vec{k}}^{(1)})' + \cH\hat{\phi}_{\vec{k}}^{(1)} 
= -4\pi Ga^2\bar{\rho}_\gamma^{(0)}(1 + w)\hat{v}_{\vec{k}}^{(1)}
~\label{e0i},\\
&&\hat{\cR}_{\vec{k}}^{(1)} - \hat{\phi}_{\vec{k}}^{(1)} = 0
~\label{eij},
\end{eqnarray}
and the evolution equation of the radiation is
\begin{equation}
(\hat{\delta}_{\vec{k}}^{(1)})' 
+ 3\cH\left( c_s^2 - w \right)\hat{\delta}_{\vec{k}}^{(1)} 
= (1 + w)\left(k^2\hat{v}_{\vec{k}}^{(1)} 
+ 3 (\hat{\cR}_{\vec{k}}^{(1)})'\right).\label{evol}
\end{equation}
Here the sound velocity, $c_s$, and equation of state, $w$, are
given by $c_s^2=w=1/3$.
Metric perturbations are represented by the one variable 
$\hat{\psi}_{\vec{k}}^{(1)} := \hat{\cR}_{\vec{k}}^{(1)} =
\hat{\phi}_{\vec{k}}^{(1)}$ from (\ref{eij}).
We substitute this into eqs.(\ref{e00}), (\ref{e0i}) and (\ref{evol})
and derive the differential equation in terms of $\hat{\psi}_{\vec{k}}^{(1)}$,
\begin{equation}
(\hat{\psi}_{\vec{k}}^{(1)})'' + \frac{4}{\eta}(\hat{\psi}_{\vec{k}}^{(1)})'
 +\frac{1}{3}k^2\hat{\psi}_{\vec{k}}^{(1)} = 0,
\end{equation}
where we used Friedmann equations and $\cH=1/\eta$ for
radiation-dominated era.
We define $y := k\eta/\sqrt{3}$ and solve this equation to obtain
\begin{equation}
\hat{\psi}^{(1)}(k,\eta) = 3\hat{\psi}_0^{(1)}(\vec{k})\frac{j_1(y)}{y},
\label{psi1}
\end{equation}
where $\hat{\psi}_0^{(1)}$ is the primordial adiabatic fluctuation and 
$j_1(y)$ is spherical Bessel's function of order 1, 
\begin{equation}
j_1(y)=\frac{1}{y^2}\left(\sin y-y\cos y\right).
\end{equation}
In solving the evolution of the fluctuations, we used eq.(\ref{evol}).
At small scale, however, we need to solve the Boltzmann equation 
with anisotropic stress.
Then the fluctuations dissipate by the Silk damping.
The dissipating fluctuation can not generate the magnetic fields.
So we consider effectively this influences by introducing the cut-off scale
later.

From eqs.(\ref{e00}), (\ref{e0i}) and (\ref{eij}),  the perturbation 
$\hat{\delta V}^{(1)}_{\vec{k}}$ is
\begin{equation}
\hat{\delta V}^{(1)}_{\vec{k}} 
= \cH \hat{v}_{\vec{k}} -\frac{1}{4}\hat{\delta}_{\vec{k}}
= \frac{1}{6\cH^2}k^2\hat{\psi}_{\vec{k}}
= \frac{1}{6}\eta^2k^2\hat{\psi}_{\vec{k}}.
\label{sk1}
\end{equation}
Using the above equation, we can write the source, $\hat{S}^i$, as
\begin{equation}
\hat{S}^i =
 \int d^3p~(\vec{p}\times\vec{k})^i
 \hat{C}_{\vec{p}}^{(1)}\hat{\psi}_{0,\vec{k}-\vec{p}}^{(1)}
 f(\vec{k},\vec{p},\eta),
\label{hs2}
\end{equation}
where 
\begin{equation}
f(\vec{k},\vec{p},\eta):=
\frac{1}{(2\pi)^{3/2}}\frac{\eta^2\bar R^{(0)}}{2(1+\bar R^{(0)})}
 \frac{1}{\bar \alpha^{(0)}}|\vec{k}-\vec{p}|^2
 \frac{j_1(y)}{y},
\end{equation}
with $y=|\vec{k}-\vec{p}|\eta/\sqrt{3}$.
Furthermore the source, $\hat{\Omega}^i$, can be written as
\begin{equation}
\hat{\Omega}^i =
 \int d^3p~(\vec{p}\times\vec{k})^i
 \hat{C}_{\vec{p}}^{(1)}\hat{\psi}_{0,\vec{k}-\vec{p}}^{(1)}
 g(\vec{k},\vec{p},\eta),
\label{hs3}
\end{equation}
where
\begin{equation}
g(\vec{k},\vec{p},\eta):=
\frac{1}{(2\pi)^{3/2}}\frac{1}{\bar \alpha^{(0)}}
\frac{|\vec{k}-\vec{p}|^2}{2\eta(1+\bar R^{(0)})}
 \int d\tilde{\eta}\frac{(\tilde{\eta})^2\bar{R}^{(0)}}{1+\bar R^{(0)}}
\frac{j_1(\tilde{y})}{\tilde{y}},
\end{equation}
with $\tilde{y}=|\vec{k}-\vec{p}|\tilde{\eta}/\sqrt{3}$.

\subsection{Power spectrum}
The power spectrum of the generated magnetic fields $P_B$ is defined as
\begin{equation}
\left<B_i(\vec{k},\eta)B^{*i}(\vec{K},\eta)\right>:=
\frac{2\pi^2}{k^3}P_B(k)\delta(\vec{k}-\vec{K}),
\end{equation}
where bracket means to take the ensemble average.
From eq.(\ref{magmag}), 
the left-hand side in the above equation is
\begin{eqnarray}
&&\left<\hat B_i(\vec{k},\eta) \hat B^{*i}(\vec{K},\eta)\right>=
\left(\frac{1-\beta^3}{1+\beta}\frac{4\sigma_T\bar\rho_{\gamma0}^{(0)}}
 {3e a^3}\right)^2\int \frac{d\eta_1}{a^2(\eta_1)}\int \frac{d\eta_2}{a^2(\eta_2)}
\times\nonumber\\&& \qquad\qquad\qquad\times
 \left(\left<\hat{\Omega}_i(\vec{k},\eta_1)\hat{\Omega}^{*i}(\vec{K},\eta_2)\right>
 + \left<\hat{S}_i(\vec{k},\eta_1)\hat{S}^{*i}(\vec{K},\eta_2)\right>\right),
\end{eqnarray}
where we neglect correlation between $\hat S$ and $\hat \Omega$, for example 
 $\left<\hat{S}\hat\Omega\right>$ for simplicity.

We estimate 
$\left<\hat{S}_i(\vec{k},\eta_1)\hat{S}^{*i}(\vec{K},\eta_2)\right>$:
\begin{eqnarray}
&&\left<\hat{S}_i(\vec{k},\eta_1)\hat{S}^{*i}(\vec{K},\eta_2)\right>
\nonumber\\ &&\qquad\qquad=
\int d^3p_1 \int d^3p_2 (\vec{k}\times\vec{p_1})_i 
(\vec{K}\times\vec{p_2})^i f(\vec{k},\vec{p}_1,\eta_1)
f(\vec{K},\vec{p}_2,\eta_2)
\times\nonumber\\&&\qquad\qquad\qquad \times
\left<\hat{C}^{(1)}(\vec{p_1})\hat{\psi}_0^{(1)}(\vec{k}-\vec{p_1})
\hat{C}^{*(1)}(\vec{p_2})\hat{\psi}_0^{*(1)}(\vec{K}-\vec{p_2})\right>
\nonumber\\
&&\qquad\qquad=
\int d^3p_1 \int d^3p_2 (\vec{k}\times\vec{p_1})_i 
(\vec{K}\times\vec{p_2})^i f(\vec{k},\vec{p}_1,\eta_1)
f(\vec{K},\vec{p}_2,\eta_2)
\times \nonumber\\&&\qquad\qquad\quad\times
\left(\left<\hat{C}^{(1)}(\vec{p_1})\hat{\psi}_0^{(1)}(\vec{k}-\vec{p_1})\right>
\left<\hat{C}^{*(1)}(\vec{p_2})\hat{\psi}_0^{*(1)}(\vec{K}-\vec{p_2})\right>
\right.\nonumber\\&&\qquad\qquad\qquad\quad
+
\left<\hat{C}^{(1)}(\vec{p_1})\hat{C}^{*(1)}(\vec{p_2})\right>
\left<\hat{\psi}_0^{(1)}(\vec{k}-\vec{p_1})\hat{\psi}_0^{*(1)}(\vec{K}-\vec{p_2})\right>
\nonumber\\&&\qquad\qquad\qquad\quad \left.
+
\left<\hat{C}^{(1)}(\vec{p_1})\hat{\psi}_0^{*(1)}(\vec{K}-\vec{p_2})\right>
\left<\hat{\psi}_0^{(1)}(\vec{k}-\vec{p_1})\hat{C}^{*(1)}(\vec{p_2})\right>\right)
\label{SS}
\end{eqnarray}
where we used Wick's theorem in the second equality.
The power spectrum of the primordial adiabatic fluctuations and non-adiabatic
fluctuations is defined respectively as
\begin{equation}
 \left<\hat \psi_0(\vec{k}) \hat \psi_0^*(\vec{K})\right> 
 := 2\pi^2P_a(k)
 \delta(\vec{k}-\vec{K})\label{pspsi}
\end{equation}
and 
\begin{equation}
 \left<\hat C(\vec{k}) \hat C^*(\vec{K})\right> 
 := 2\pi^2P_{na}(k)
 \delta(\vec{k}-\vec{K}).
\end{equation}
We assume no correlation between adiabatic and non-adiabatic mode,
$\left<\hat C \hat \psi_0\right>=0$.
Hence eq.(\ref{SS}) becomes 
\begin{eqnarray}
\left<\hat{S}_i(\vec{k},\eta_1)\hat{S}^{*i}(\vec{K},\eta_2)\right>
&=&
(2\pi^2)^2\delta(\vec{k}-\vec{K}) \int d^3p (\vec{k}\times\vec{p})_i 
(\vec{K}\times\vec{p})^i \times
\nonumber\\&&\times
P_{na}(p)P_a(|\vec{k}-\vec{p}|)
f(\vec{k},\vec{p},\eta_1)
f(\vec{K},\vec{p},\eta_2)
\end{eqnarray}
and similarly
\begin{eqnarray}
\left<\hat{\Omega}_i(\vec{k},\eta_1)\hat{\Omega}^{*i}(\vec{K},\eta_2)\right>
&=&
(2\pi^2)^2\delta(\vec{k}-\vec{K}) \int d^3p (\vec{k}\times\vec{p})_i 
(\vec{K}\times\vec{p})^i 
\nonumber\\&&\times
P_{na}(p)P_a(|\vec{k}-\vec{p}|)
g(\vec{k},\vec{p},\eta_1)
g(\vec{K},\vec{p},\eta_2).
\end{eqnarray}
Finally we obtain the power spectrum of the magnetic fields as
\begin{eqnarray}
&&\frac{2\pi^2}{k^3}P_B(k)
=
\left(\frac{1-\beta^3}{1+\beta}\frac{\sigma_T\bar \rho^{(0)}_{\gamma0}}
 {e a^3}\right)^2
(2\pi^2)^2 \int d^3p |\vec{k}\times\vec{p}|^2 
 P_{na}(p)P_a(|\vec{k}-\vec{p}|)
\times\nonumber\\&&\times
\int^{\eta}_0 d\eta_1\int^{\eta}_0 d\eta_2
a_1^{-2}a_2^{-2}
\{g(\vec{k},\vec{p},\eta_1)g(\vec{k},\vec{p},\eta_2)
+f(\vec{k},\vec{p},\eta_1)f(\vec{k},\vec{p},\eta_2)\},
\label{P_B}
\end{eqnarray}
where
\begin{eqnarray}
&&f(\vec{k},\vec{p},\eta):=
\frac{1}{(2\pi)^{3/2}}\frac{\eta^2\bar R^{(0)}}{2(1+\bar R^{(0)})}
 \frac{1}{\bar \alpha^{(0)}}|\vec{k}-\vec{p}|^2
 \frac{j_1(y)}{y},
\\
&&g(\vec{k},\vec{p},\eta):=
\frac{1}{(2\pi)^{3/2}}\frac{1}{\bar \alpha^{(0)}}
\frac{|\vec{k}-\vec{p}|^2}{2\eta(1+\bar R^{(0)})}
 \int d\tilde{\eta}\frac{(\tilde{\eta})^2\bar{R}^{(0)}}{1+\bar R^{(0)}}
\frac{j_1(\tilde{y})}{\tilde{y}},
\\
&& \bar \alpha^{(0)} 
  = \frac{(1 + \beta^2)(1 + \bar R^{(0)})}{1 + \beta}
  \frac{a\sigma_T \bar\rho_\gamma^{(0)}}{m_p},
\end{eqnarray}
$\bar R ^{(0)}=(3m_p(1+\beta)\bar n^{(0)})/(4\bar\rho^{(0)}_\gamma)$, 
and $y=|\vec{k}-\vec{p}|\eta/\sqrt{3}$.

In the end of this section, we note about the power spectra of the
adiabatic and non-adiabatic fluctuations.
The power spectrum of the adiabatic mode, $P_a(k)$, can be well
described as a power law
\begin{equation}
P_a(k)=\frac{1}{k^3}\Delta^2_{\cR}(k_0)\left(\frac{k}{k_0}\right)^{n_1-1}~,
\end{equation}
where $k_0=0.002 {\rm Mpc^{-1}}$ is the pivot scale,
$\Delta_{\cR}^2\simeq 2.5\times 10^{-9}$, and $n_1\simeq 1.0$ 
from the 7-year WMAP data \cite{Komatsu:2010fb}.  
We assume that the non-adiabatic fluctuations have also a power law.
The power spectrum of the non-adiabatic fluctuations of baryon is
\begin{equation}
\frac{P_{na}}{P_a} 
:= \left(\frac{\Omega_{\rm CDM}}{\Omega_b}\right)^2\frac{\cA}{1-\cA}\left(\frac{k}{k_0}\right)^{n_2-1},
\end{equation}
where $\cA$ is represented as the ratio of the cold dark matter (CDM) isocurvature
fluctuation to the sum of the adiabatic and the 
isocurvature fluctuations. 
The factor of $\left(\Omega_{\rm CDM}/\Omega_b\right)^2$
comes in order to convert the power spectrum of CDM isocurvature mode
to that of the baryon isocurvature one $C(\vec{x})$ \cite{Bucher:1999re}.
Recent observations indicate a blue spectrum with $n_2=2-4$ and 
$\cA \lesssim  0.1-0.01$, depending on the observational data used
\cite{Komatsu:2010fb,Sollom:2009vd,Li:2010yb}.
In this paper, we adopt $\cA\simeq 0.01$ and $n_2=4$ as a simple
example.
Since the amplitude of the generated magnetic fields is proportional 
to the root of the amplitude of the non-adiabatic fluctuations,
$B\propto\sqrt{P_B}\propto\sqrt{\cA}$, it is insentive 
for  the amplitude of the non-adiabatic fluctuations.
Therefore the amplitude of the non-adiabatic fluctuation is less
important.
We give the power spectrum of the generated magnetic fields 
for the different index of the non-adiabatic fluctuations later.

\section{Results}
\subsection{Integral region and Results}

\begin{figure}[t]
\begin{center}
\includegraphics[width=0.5\textwidth]{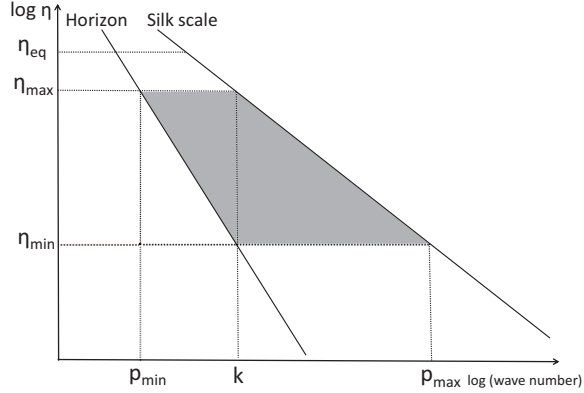}
\caption{The integral region of eq.(\ref{P_B}).
$\eta_{eq}$ is the conformal time which is the radiation-matter 
equality time. The horizon is depend on the time, such as
 $k_{H}\propto\eta^{-1}$ and the Silk scale is also depend on the time,
 such as $k_{Silk}\propto\eta^{-3/2}$. 
Then the maximum and minimum of the wavenumber and the conformal time
 are determined for an arbitrary wavenumber, $k$, of the magnetic fields.}
\label{IR}
\end{center}
\end{figure}

We need to perform numerical integrations of eq.(\ref{P_B}).
As a cut-off scale we consider the Silk dumping which is the dissipation 
of fluctuations by a radiation drag. The fluctuations whose scale is
less than the Silk scale dissipate and do not create the magnetic
fields any more.

We take it into account that the Silk scale depends on time
$k_{Silk}\simeq 4.3\times10^{-6}a^{-3/2} {\rm Mpc}^{-1}$
\cite{Ichiki:2007hu,Hu}.
Furthermore we neglect the fluctuations with larger scale at each time 
than the horizon because large-scale modes do not contribute 
to the generation of the magnetic fields.
The integral region for a wave number $k$ is presented as the gray 
trapezoid region in Fig. \ref{IR}.

\begin{figure}[t]
\begin{center}
\includegraphics[width=0.65\textwidth]{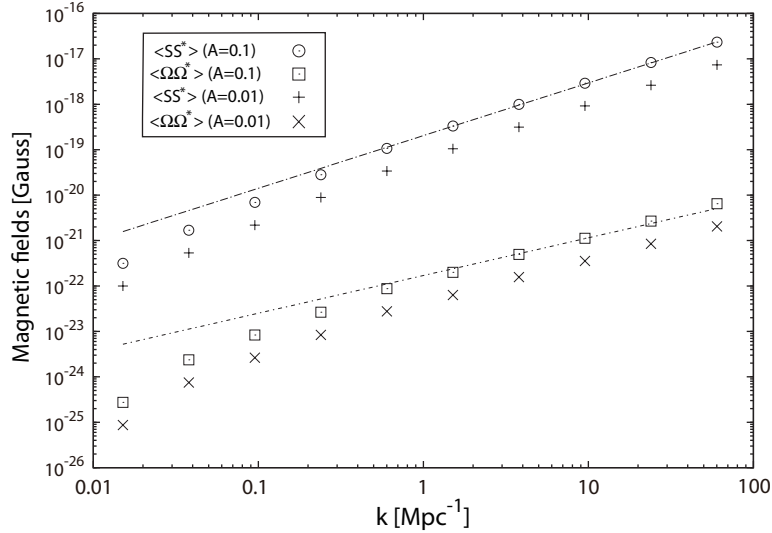}
\caption{Calculated physical amplitudes of magnetic fields
for several comoving wave numbers ($k$). The plus mark and
 cross mark represent the contributions of the slip term and
vorticity term respectively. Here we set $\cA\simeq 0.1$(circle and
 square mark), $\cA\simeq 0.01$(plus and cross mark) and $n_2=4$. 
Two lines are the analytically-estimated power spectra (see text).}
\label{figmag}
\end{center}
\end{figure}

We perform the numerical integration for $\mathcal A \simeq 0.1,0.01$ 
and $n_2\simeq 4$ to calculate the amplitudes of the magnetic fields
at the radiation-matter equality time and show the result in Fig. \ref{figmag}.
The circle and plus mark represent the contribution of the slip term 
$\left<SS^{*}\right>$ for each $\cA$ and
the square and cross mark represent the contribution of the vorticity term 
$\left<\Omega\Omega^{*}\right>$ for each $\cA$.
From this figure we find that the vorticity term is less important at
the generation of the magnetic fields at pre-recombination era. 
The physical magnetic field strength is about $10^{-20}$ Gauss at
1Mpc.
The generated magnetic fields have a blue spectrum $k^{1.16}$ and so 
the larger strength is expected at the smaller scales.


\subsection{Simple estimation for arbitrary spectral index $n_2$}
Finally we show a simple estimation of the k-dependence of the magnetic-field
spectra.
For the slip term, the power spectrum of the magnetic fields is 
$P_B\sim k^5p^{n_2+1}|\vec{k}-\vec{p}|^{-3}(\int f\eta^{-2} d\eta)^2 $.
Since $\int \eta^{-2}f d\eta\sim \eta^2|\vec{k}-\vec{p}|^{-1}$,
$P_B\sim k^5p^5\eta^4/|\vec{k}-\vec{p}|^5 $.
Since the main contribution to the integration for $p$ 
is from $p\sim |\vec{k}-\vec{p}|\sim k$, we
 obtain $P_B\sim k^{n_2+1}\eta^4 $.
We consider the silk scale $\eta\propto k^{-2/3}$ and obtain 
$B\sim \sqrt{P_B}\sim k^{n_2/2-5/6} $ as a behavior of the magnetic fields.
For $n_2=4$, $B\sim k^{7/6 \sim 1.16}$.
We take the same process about vorticity term and obtain 
$B\sim \sqrt{P_B}\sim k^{n_2/2-7/6}$.
For $n_2=4$, $B\sim k^{5/6 \sim 0.83}$.
The deviation from this power law at the large scale is caused 
by the boundary effect from the horizon.
This fact indicates that the contribution at the Silk scale is very
important in the generation of the fields at small scales.

\section{Conclusion}
In this paper, we discussed the generation of magnetic fields
by non-adiabatic fluctuations at the pre-recombination era.
Firstly we showed analytically that non-adiabatic fluctuations generate
magnetic fields using the tight coupling approximation.  We found
that magnetic fields are generated at the first-order
expansion of the tight coupling approximation. This result should be
compared with the case of initially-adiabatic fluctuations, where
magnetic fields are created only at the second order.

Secondly we calculated the power spectra of magnetic fields
at the radiation-matter equality time considering the non-adiabatic
fluctuations with a blue spectrum.  For this we only consider the second
order coupling between primordial non-adiabatic and adiabatic
fluctuations because this coupling is expected to give larger magnetic fields
than the auto-coupling of non-adiabatic fluctuations.
We found that the fields have a blue
spectrum $B\propto k^{1.16}$ for $n_2=4$, and the amplitude is about
$B\sim10^{-20}$ Gauss at the comoving scale of 1Mpc if the maximum amount
of non-adiabatic fluctuations allowed from CMB observations is considered.
Because the spectrum is bluer than those from the adiabatic fluctuations
found in refs. \cite{Ichiki:2007hu,Fenu:2010kh}, this fields may dominate
the primordial magnetic fields at small scales.
The amplitude would be enough to be amplified to the present
fields in the galaxy by dynamo mechanism \cite{Davis:1999bt}, however
it is still insufficient to explain the fields in void regions recently
claimed by
\cite{Ando:2010rb,Essey:2010nd,Neronov:1900zz,Dolag:2010ni,Takahashi:2011ac}.  
Clearly both detailed observations and theoretical investigations are
necessary to understand how the large-scale magnetic fields
evolve after their generation in voids.

Because we consider only the first order of the tight coupling approximation,
we cannot directly extend our analysis to the time of recombination.
For a complete analysis, we have to solve the time evolution
of $\Delta v_{(\gamma b)}$ numerically.  This will be our future work.

\section*{Acknowledgments}
S.M. is supported by JSPS Grant-Aid for Scientific Research (No.10J00547).
This work has been supported in part by Grant-in-Aid for Scientific
Research Nos. 21740177, 22012004 (K.I.), 21840028 (K.T.) of the Ministry of
Education, Sports, Science and Technology (MEXT) of Japan, and also
supported by Grant-in-Aid for the Global Center of Excellence program at
Nagoya University "Quest for Fundamental Principles in the Universe:
from Particles to the Solar System and the Cosmos" from the MEXT of
Japan.


\appendix
\section{Four velocities, vorticities and electromagnetic fields}
In this appendix, we write down four velocities, vorticities and
electromagnetic fields used in our paper.
From the metric and normalized condition $u_\mu u^\mu=-1$, 
the four vector of the each species $u_{(I),\mu}$ is
\begin{eqnarray}
 u_{(I)\mu} &=& 
     a\Bigl(-1-\phi+\frac{1}{2}\phi^2-\frac{1}{2}v_{(I)j}v_{(I)}^j, 
v_{(I)i}+\chi_i-2\cR v_{(I)i} \Bigr),
\label{four2}
\end{eqnarray}
where $v_{(I),i}$ is the 3-dimensional velocity.
The energy-momentum tensor of each spices($I=p,e,\gamma$) is represented
as 
\begin{equation}
T_{(I)\mu}^{\;\;\;\nu} = (\rho_I +p_I)u_{(I)\mu} u_{(I)}^\nu + 
p_I\delta_\mu^{~\nu} 
~.
\end{equation}
where $p_\gamma=\rho_\gamma/3$ and $p_{p,e}=0$.
We give the useful relation
\begin{eqnarray}
\nabla_\nu T_{(I)0}{}^{\nu}
&=&  -\left[\rho_I' +3\cH(\rho_I +p_I)\right]
     +3(\rho_I+p_I)\cR' 
\nonumber\\& &\qquad\qquad
-(\rho_I+p_I)\dd_i v_{(I)}^i, \label{En1}
\\
\nabla_\nu T_{(I)i}{}^{\nu}
&=&   \left[(\rho_I+p_I)(v_{(I)i}+\chi_i)\right]' +4\cH(\rho_I+p_I)(v_{(I)i}+\chi_i)
\nonumber\\&&\quad
+(\rho_I+p_I)\dd_i(\phi-\phi^2) +\dd_i p_I
\nonumber \\&&
 -\left(\phi+2\cR\right)\left\{ \left[\left(\rho_I+p_I\right) v_{(I)i}\right]'
 +4\cH\left(\rho_I+p_I\right) v_{(I)i}\right\}
\nonumber\\&&\quad\qquad
 -5(\rho_I+p_I)\cR'v_{(I)i} +(\rho_I+p_I)\dd_j\left(v_{(I)i}v_{(I)}^j\right),
\end{eqnarray}

The vorticity of photons is defined as 
$\omega_{\gamma}^i:=-\frac{1}{2}\varepsilon^{i\nu\rho\lambda}u_{(\gamma)\lambda}\nabla_\nu
u_{(\gamma)\rho}$ 
where $\varepsilon^{i\nu\rho\lambda}$ is  the alternative tensor with  $\varepsilon^{0123}=1/\sqrt{-g}$.
Since the first-order vector perturbation does not exist,
the leading term of the vorticity is second-order perturbation.
\begin{equation}
\omega_\gamma^{(2)i} = -\frac{1}{2a^2}\epsilon^{ijk}
                      \left[\dd_j\left(v_{(\gamma)k}^{(2)} + \chi_{k}^{(2)}\right)
    + v_{(\gamma)j}^{(1)}v_{(\gamma)k}^{(1)'} +v_{(\gamma)j}^{(1)}
\dd_k\left(\phi^{(1)}+2\cR^{(1)}\right)\right],
\label{appom}
\end{equation}
where $\epsilon^{ijk}$ is the 3-dimensional flat alternating tensor with
$\epsilon^{123}=1$.

The Faraday tensor $F_{\mu\nu}$ is decomposed to the electric 
and magnetic fields
\begin{equation}
E^\mu = F^{\mu\nu}u_{(\gamma) \nu} , \quad
B^\mu = \frac{1}{2}\varepsilon^{\mu\nu\lambda\sigma}u_{(\gamma)\sigma}
F_{\nu\lambda}~.
\label{appind}
\end{equation}


\section*{References}
\begin{thebibliography}{99}
\bibitem{Grasso:2000wj}
  D.~Grasso and H.~R.~Rubinstein,
  Phys.\ Rept.\  {\bf 348}, 163 (2001)
  [arXiv:astro-ph/0009061].

\bibitem{Widrow:2002ud}
  L.~M.~Widrow,
  Rev.\ Mod.\ Phys.\  {\bf 74}, 775 (2002)
  [arXiv:astro-ph/0207240].

\bibitem{Giovannini:2003yn}
  M.~Giovannini,
  Int.\ J.\ Mod.\ Phys.\  D {\bf 13}, 391 (2004)
  [arXiv:astro-ph/0312614].

\bibitem{Kandus:2010nw}
  A.~Kandus, K.~E.~Kunze, C.~G.~Tsagas,
  Phys.\ Rept.\  {\bf 505 } (2011)  1-58.
  [arXiv:1007.3891 [astro-ph.CO]].



\bibitem{Yamazaki:2008gr}
  D.~G.~Yamazaki, K.~Ichiki, T.~Kajino and G.~J.~Mathews,
  Phys.\ Rev.\  D {\bf 77}, 043005 (2008)
  [arXiv:0801.2572 [astro-ph]].

\bibitem{Yamazaki:2010nf}
  D.~G.~Yamazaki, K.~Ichiki, T.~Kajino and G.~J.~Mathews,
  Phys.\ Rev.\  D {\bf 81}, 023008 (2010)
  [arXiv:1001.2012 [astro-ph.CO]].

\bibitem{Kahniashvili:2010wm}
  T.~Kahniashvili, A.~G.~Tevzadze, S.~K.~Sethi, K.~Pandey and B.~Ratra,
  Phys.\ Rev.\  D {\bf 82}, 083005 (2010)
  [arXiv:1009.2094 [astro-ph.CO]].

\bibitem{Tashiro:2009hx}
  H.~Tashiro and N.~Sugiyama,
  Mon. Not. Roy. Astron. Soc. {\bf 411}, 1284 (2011)
  arXiv:0908.0113 [astro-ph.CO].

\bibitem{Paoletti:2010rx}
  D.~Paoletti, F.~Finelli,
  Phys.\ Rev.\  {\bf D83 } (2011)  123533.
  [arXiv:1005.0148 [astro-ph.CO]].

\bibitem{Schleicher:2011jj}
  D.~R.~G.~Schleicher and F.~Miniati,
  arXiv:1108.1874 [astro-ph.CO].




\bibitem{Ando:2010rb}
  S.~Ando and A.~Kusenko,
  Astrophys.\ J.\  {\bf 722}, L39 (2010)
  [arXiv:1005.1924 [astro-ph.HE]].

\bibitem{Essey:2010nd}
  W.~Essey, S.~Ando and A.~Kusenko,
  arXiv:1012.5313 [astro-ph.HE].

\bibitem{Neronov:1900zz}
  A.~Neronov and I.~Vovk,
  Science {\bf 328} (2010) 73
  [arXiv:1006.3504 [astro-ph.HE]].

\bibitem{Dolag:2010ni}
  K.~Dolag, M.~Kachelriess, S.~Ostapchenko and R.~Tomas,
  Astrophys.\ J.\  {\bf 727}, L4 (2011)
  [arXiv:1009.1782 [astro-ph.HE]].

\bibitem{Takahashi:2011ac}
  K.~Takahashi, M.~Mori, K.~Ichiki and S.~Inoue,
  arXiv:1103.3835 [astro-ph.CO].


\bibitem{Berezhiani:2003ik}
  Z.~Berezhiani, A.~D.~Dolgov,
  Astropart.\ Phys.\  {\bf 21 } (2004)  59-69.
  [astro-ph/0305595].

\bibitem{Matarrese:2004kq}
  S.~Matarrese, S.~Mollerach, A.~Notari and A.~Riotto,
  Phys.\ Rev.\  D {\bf 71}, 043502 (2005)
  [arXiv:astro-ph/0410687].

\bibitem{Takahashi:2005nd}
  K.~Takahashi, K.~Ichiki, H.~Ohno and H.~Hanayama,
  Phys.\ Rev.\ Lett.\  {\bf 95} (2005) 121301
  [arXiv:astro-ph/0502283].

\bibitem{Kobayashi:2007wd}
  T.~Kobayashi, R.~Maartens, T.~Shiromizu and K.~Takahashi,
  Phys.\ Rev.\  D {\bf 75}, 103501 (2007)
  [arXiv:astro-ph/0701596].

\bibitem{Ichiki:2007hu}
  K.~Ichiki, K.~Takahashi, N.~Sugiyama, H.~Hanayama and H.~Ohno,
  arXiv:astro-ph/0701329.

\bibitem{Maeda:2008dv}
  S.~Maeda, S.~Kitagawa, T.~Kobayashi and T.~Shiromizu,
  Class.\ Quant.\ Grav.\  {\bf 26}, 135014 (2009)
  [arXiv:0805.0169 [astro-ph]].

\bibitem{Fenu:2010kh}
  E.~Fenu, C.~Pitrou and R.~Maartens,
  arXiv:1012.2958 [astro-ph.CO].



\bibitem{Bassett:2005xm}
  B.~A.~Bassett, S.~Tsujikawa and D.~Wands,
  Rev.\ Mod.\ Phys.\  {\bf 78}, 537 (2006)
  [arXiv:astro-ph/0507632].



\bibitem{Sollom:2009vd}
  I.~Sollom, A.~Challinor and M.~P.~Hobson,
  Phys.\ Rev.\  D {\bf 79}, 123521 (2009)
  [arXiv:0903.5257 [astro-ph.CO]].

\bibitem{Li:2010yb}
  H.~Li, J.~Liu, J.~Q.~Xia and Y.~F.~Cai,
  arXiv:1012.2511 [astro-ph.CO].


\bibitem{Christopherson:2009bt}
  A.~J.~Christopherson, K.~A.~Malik and D.~R.~Matravers,
  Phys.\ Rev.\  D {\bf 79}, 123523 (2009)
  [arXiv:0904.0940 [astro-ph.CO]].

\bibitem{Christopherson:2010ek}
  A.~J.~Christopherson, K.~A.~Malik and D.~R.~Matravers,
  arXiv:1008.4866 [astro-ph.CO].


\bibitem{Brandenberger:1998ew}
  R.~H.~Brandenberger, X.~-m.~Zhang,
  Phys.\ Rev.\  {\bf D59 } (1999)  081301.
  [hep-ph/9808306].

\bibitem{Gwyn:2008fe}
  R.~Gwyn, S.~H.~Alexander, R.~H.~Brandenberger, K.~Dasgupta,
  Phys.\ Rev.\  {\bf D79 } (2009)  083502.
  [arXiv:0811.1993 [hep-th]].



\bibitem{Pitrou:2010ai}
  C.~Pitrou,
  Phys.\ Lett.\  B {\bf 698}, 1 (2011)
  [arXiv:1012.0546 [astro-ph.CO]].


\bibitem{Komatsu:2010fb}
  E.~Komatsu {\it et al.}  [WMAP Collaboration],
  Astrophys.\ J.\ Suppl.\  {\bf 192}, 18 (2011)
  [arXiv:1001.4538 [astro-ph.CO]].



\bibitem{Bartolo:2006cu}
  N.~Bartolo, S.~Matarrese and A.~Riotto,
  JCAP {\bf 0606}, 024 (2006)
  [arXiv:astro-ph/0604416].

\bibitem{Pitrou:2008ut}
  C.~Pitrou,
  Gen.\ Rel.\ Grav.\  {\bf 41}, 2587 (2009)
  [arXiv:0809.3245 [astro-ph]].



\bibitem{Bucher:1999re}
  M.~Bucher, K.~Moodley, N.~Turok,
  Phys.\ Rev.\  {\bf D62}, 083508 (2000)
  [astro-ph/9904231].

\bibitem{Hu}
  W.~Hu, N.~Sugiyama,
  Astrophys.\ J.\ {\bf 444}, 489 (1995)
  [arXiv:astro-ph/9510117].

\bibitem{Davis:1999bt}
  A.~C.~Davis, M.~Lilley and O.~Tornkvist,
  Phys.\ Rev.\  D {\bf 60}, 021301 (1999)
  [arXiv:astro-ph/9904022].



\end{thebibliography}
\end{document}